# Correspondence on "Controlling the Curie temperature in (Ga,Mn)As through location of the Fermi level within the impurity band" by Dobrowolska *et al.*

K. W. Edmonds[1], B. L. Gallagher[1], M. Wang[1], A. W. Rushforth[1], O. Makarovsky[1], A. Patanè[1], R. P. Campion[1], C. T. Foxon[1], V. Novák[2], and T. Jungwirth[2,1]

[1] *School of Physics and Astronomy, University of Nottingham, Nottingham NG7 2RD, United Kingdom*

[2] *Institute of Physics ASCR, v.v.i., Cukrovanická 10, 16253 Praha 6, Czech Republic*

(Ga,Mn)As has become a prototypical model system for semiconductor spintronics research. It is therefore important to understand and to control its semiconducting and magnetic properties. Since the discovery of ferromagnetism in (Ga,Mn)As in the 1990s, a key focus has been the development of a microscopic understanding of the factors determining the Curie temperature $T_C$ and on this basis developing strategies for raising $T_C$. The main statements and conclusions of the recent Nature Materials article by Dobrowolska *et al.*[1] on this topic can be summarized as follows: (i) Experimental doping trends for $T_C$ in (Ga,Mn)As are inconsistent with calculations in Ref. 2 which are based on the valence band model of (Ga,Mn)As. (ii) Experimental doping trends are consistent with an impurity band model of (Ga,Mn)As. (iii) The results open new avenues for achieving higher values of $T_C$. (iv) Conclusions (i)-(iii) are possible because the ion channeling experiments presented give reliable effective local moment and hole densities. Here we argue that all the above points are incorrect.

(i) The calculations in Ref. 2 are not based on a "valence band model". These are microscopic multi-orbital tight-binding Anderson calculations in which the merged valence and impurity bands in ferromagnetic (Ga,Mn)As are a result of the calculations, not a model assumption. From this perspective they are conceptually analogous to *ab initio* calculations and they yield very similar band structure of (Ga,Mn)As to the full-potential GGA+U density functional theory.[3] Dobrowolska *et al.*[1] omitted to mention that Ref. 2 also contains experimental data combining high-field Hall and SQUID magnetometry measurements which show that the doping trends of $T_C$ are in broad agreement with the microscopic theory. In particular, the experimental data in Ref. 2 do not show the collapse of $T_C$ at low compensation seen in Fig. 1 of Ref. 1. For Dobrowolska *et al.*[1] this collapse seen in their data is the key argument for the failure of the theory of (Ga,Mn)As presented in Ref. 2. In Fig. 1 of this communication we provide an extended set of measurements on our (Ga,Mn)As samples which fully supports the theoretical and experimental data of Ref. 2 and which, at low compensation, completely disagrees with the measurements presented by Dobrowolska *et al.*[1]. To further highlight these points we plot in Fig. 2 the low temperature conductivities for a series of our (Ga,Mn)As films with varying Mn concentration, in which the compensation is decreased in one or more post-growth annealing steps. For a given Mn concentration, the conductivities increase with increasing hole density and are highest at close to zero compensation. In contrast, those samples reported to be close to zero compensation in Ref. 1 have very low conductivities.

(ii) The claimed agreement between the data of Ref. 1 and the impurity band model is solely based on a cartoon in which ~50% (0%) compensation corresponds to a half- (fully-) filled detached impurity band. To the best of our knowledge, no microscopic atomic orbital based theory of (Ga,Mn)As explains how the 0.1eV acceptor level of Mn could remain sufficiently narrow to avoid overlapping and mixing with the valence band at $10^{20}$-$10^{21}$ cm$^{-3}$ doping levels. It is striking that Ref. 1 does not present, or refer to, any calculated $T_C$ data obtained from their preferred impurity band theory. The evidence for the claimed agreement between the impurity band theory and experimental data in Ref. 1 is therefore missing.

(iii) The statement regarding new avenues for achieving higher values of $T_C$ is central in Ref. 1. However, the samples of Ref. 1 have a maximum $T_C$ of 90 K which is actually 20 degrees below the $T_C$ reported by Ohno in his seminal paper from 1998[4]. So the results of Ref. 1 have not demonstrably led to high $T_C$. In contrast, materials development guided by the expectation that the highest $T_C$ will occur close to zero compensation have led to $T_C$ values reaching 188 K[5,6,7].

(iv) We disagree with Dobrowolska *et al.*[1] that ion channeling analysis allows a direct evaluation of the local moment and hole densities: it is not a direct measurement of either quantity, and it neglects other possible compensating defects and sample inhomogeneities. For our uniform annealed thin film materials, ion channeling measurements (performed by a co-author of Ref. 1 and originally reported in Ref. 8) provide estimates of the hole density that agree with high-field

Hall measurements within the quoted uncertainties (see Table I). We emphasize that these samples show lower than 10% compensation from both channeling and Hall measurements and high $T_C$, as illustrated by the stars in Fig. 1. The collapse of $T_C$ at low compensations is absent in our samples independent of the method used to extract the doping densities. On the other hand, the presence of common compensating defects such as As antisites would not be detected by ion channeling. Concentrations of such defects can depend strongly on growth conditions[9] and thus can vary significantly from sample to sample, and this could account for the suppression of $T_C$ and conductivity in the materials of Ref. 1. In addition, material inhomogeneity has been observed in several previous experiments[10,11] by some of the co-authors of Ref. 1, but this does not appear to have been considered when interpreting the channeling data of Ref. 1.

We conclude that the discrepancy between our results and the results of Dobrowolska *et al.*[1] lies in the different material quality of the studied samples. Since our materials reach almost 100 degrees higher $T_C$'s than materials studied in Ref. 1, our samples are clearly of higher quality from the Curie temperature perspective. A detailed study co-authored by several of us presents compelling evidence that high quality materials suitable for the investigation of the intrinsic material properties of (Ga,Mn)As can only be prepared by careful optimization of growth and post-growth annealing parameters which has to be performed individually for each Mn doping concentration[7]. The data of Fig. 1 and 2, which are for samples prepared in this way, demonstrate that the $T_C$ and conductivity of these samples are *maximized* at low compensation rather than being zero and thus the magnetic order in (Ga,Mn)As is *not* consistent with the isolated impurity band scenario of Dobrowolska *et al.*[1]

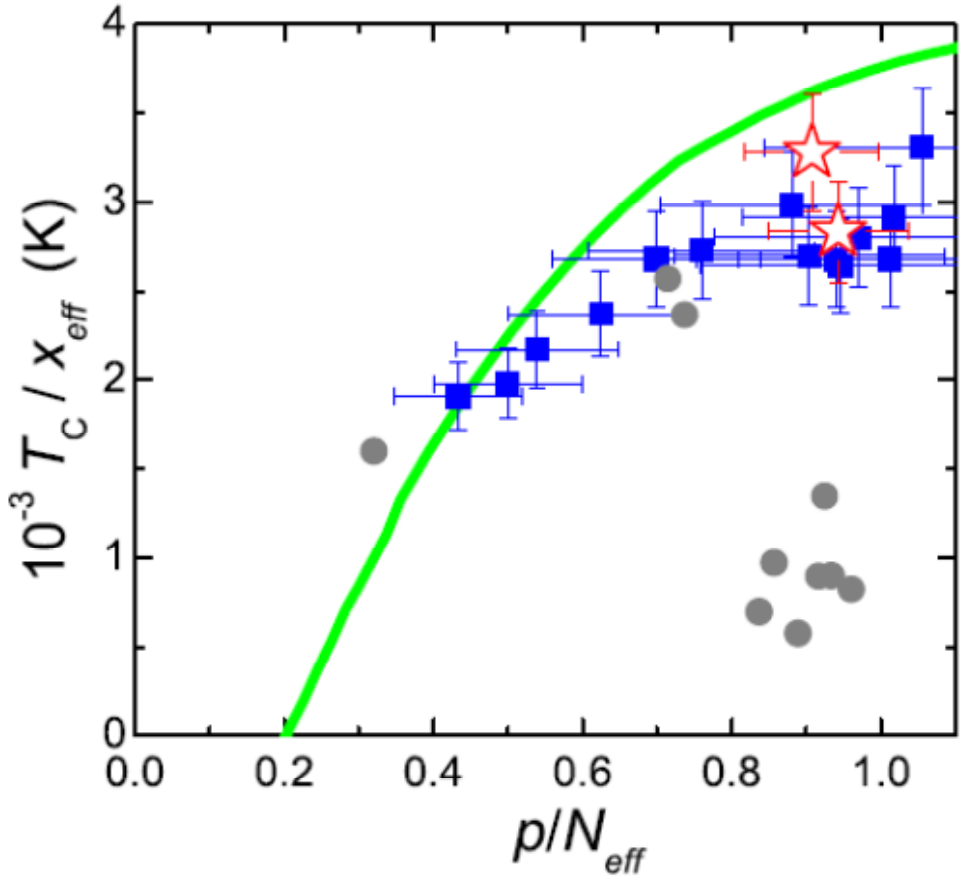

Figure 1. Comparison of $T_C/x_{eff}$ versus $p/N_{eff}$ from Ref. 2 (blue squares, hole density $p$ from high field Hall measurements), Ref. 1 (grey circles, $p$ from ion channeling measurements), and Ref. 8 (red stars, $p$ from ion channeling measurements). The green curve is the microscopic theory from Ref. 2. $x_{eff}$ and $N_{eff}$ are respectively the effective local moment doping and density, as defined in Refs. 1 and 2.

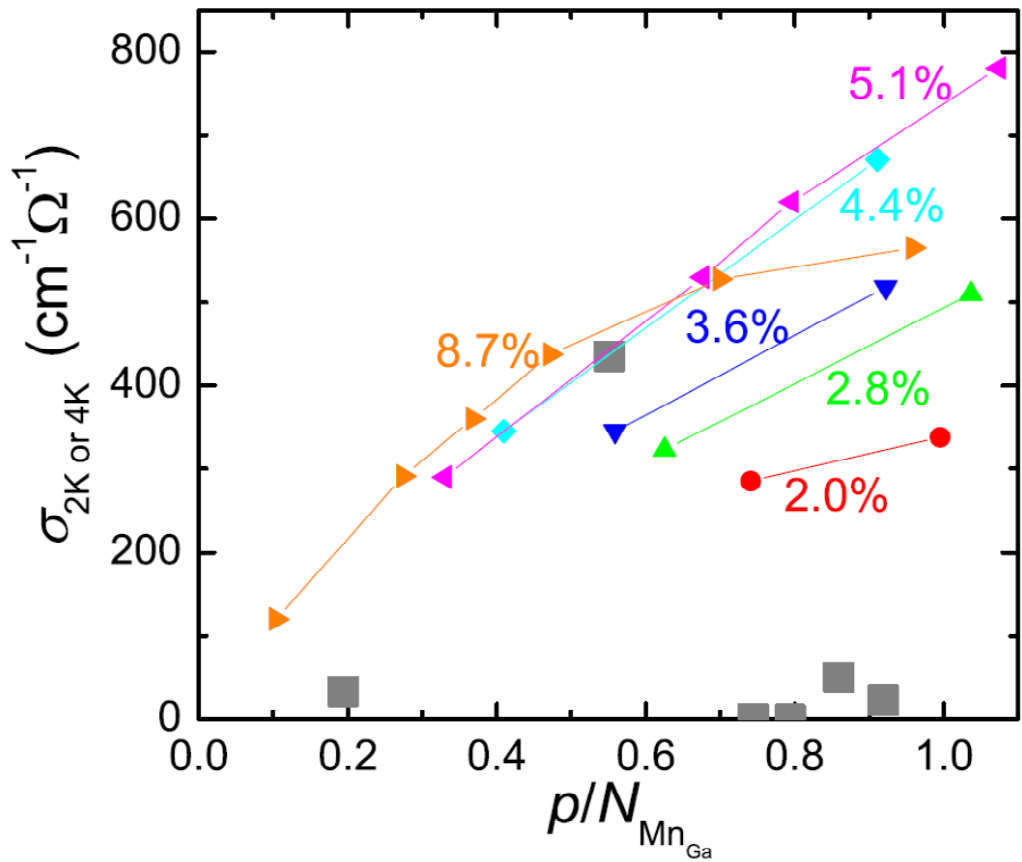

Figure 2. Low temperature conductivities versus the ratio of hole density $p$ to Mn acceptor density $N_{MnGa}$. The labels denote the value of $N_{MnGa}$ corresponding to each symbol. The grey squares are data from Ref. 1.

| Sample | Curie temperature $T_C$ (K) | Ion channeling results | | | | $p$ from high field Hall ($10^{20}$ cm$^{-3}$) |
|---|---|---|---|---|---|---|
| | | $x_{sub}$ | $x_I$ | $x_{eff}$ | $p$ ($10^{20}$ cm$^{-3}$) | |
| Ga$_{0.94}$Mn$_{0.06}$As | 128 | 0.043 | 0.004 | 0.039 | 7.8 ± 0.9 | 9.8 ± 2.0 |
| Al$_{0.1}$Ga$_{0.84}$Mn$_{0.06}$As | 119 | 0.044 | 0.002 | 0.042 | 8.7 ± 0.9 | 8.5 ± 1.7 |

Table 1. Comparison of ion channeling and Hall effect results for annealed (Ga,Mn)As and (Al,Ga,Mn)As films. The ion channeling results give the substitutional, $x_{sub}$, and interstitial, $x_I$, Mn concentrations. The hole concentration is then obtained from $p = 4(x_{sub} - 2x_I)/a^3$ ) as is done in Ref. 1. The hole concentration $p$ from high field Hall measurements have an uncertainty of around 20% (see Ref. 2). The ion channeling measurements are from Ref. 8.